\documentclass[amsmath, amssymb, preprintnumbers, showpacs,reprint, superscriptaddress,showkeys,aps,prl,twocolumn]{revtex4-1}

\usepackage{color} 
\usepackage{epsfig}
\usepackage{soul}
\usepackage{hyperref}
\usepackage{slashed}
\usepackage{graphicx}
\usepackage{amsmath}
\usepackage{latexsym}
\usepackage{epstopdf}
\usepackage{amsmath}
\usepackage{enumitem}
\usepackage{amssymb,amsmath}
\usepackage{multirow}
\usepackage{mathtools}
\usepackage{bbm}
\usepackage{bm}
\usepackage{amsfonts}
\usepackage{amssymb}
\usepackage{calligra}
\usepackage{calrsfs}
\usepackage{makecell}
\usepackage[normalem]{ulem}
\usepackage{braket}
\usepackage[british,UKenglish,USenglish,american]{babel}

\begin{document}

\title{Quantum Bubbles in Microgravity}

\author{A. Tononi}
\email{andrea.tononi@phd.unipd.it}
\affiliation{Dipartimento di Fisica e Astronomia ``Galileo Galilei'',  
Universit\`a di Padova, via Marzolo 8, Padova 35131, Italy}

\author{F. Cinti}
\affiliation{Dipartimento di Fisica e Astronomia, Universit\`a di Firenze, I-50019, Sesto Fiorentino (FI), Italy}
\email{fabio.cinti@unifi.it}
\affiliation{INFN, Sezione di Firenze, I-50019, Sesto Fiorentino (FI), Italy}
\affiliation{Department of Physics, University of Johannesburg, P.O. Box 524, Auckland Park 2006, South Africa}

\author{L. Salasnich}
\email{luca.salasnich@unipd.it}
\affiliation{Dipartimento di Fisica e Astronomia ``Galileo Galilei'',  
Universit\`a di Padova, via Marzolo 8, Padova 35131, Italy}
\affiliation{Istituto Nazionale di Ottica (INO) del Consiglio Nazionale delle 
Ricerche (CNR), \\ via Nello Carrara 1, Sesto Fiorentino 50125, Italy}

\date{\today}

\begin{abstract}
The recent developments of microgravity experiments with ultracold atoms 
have produced a relevant boost in the study of shell-shaped ellipsoidal Bose-Einstein condensates. 
For realistic bubble-trap parameters, here we calculate the critical temperature of Bose-Einstein condensation, which, if compared to the one of the {bare harmonic trap} with the same frequencies, shows a strong reduction. 
We simulate the zero-temperature density distribution with the Gross-Pitaevskii equation, and we study the free expansion of the hollow condensate. 
While part of the atoms expands in the outward direction, the condensate self-interferes inside the bubble trap, filling the hole in experimentally observable times.
For a mesoscopic number of particles in a strongly interacting regime, for which more refined approaches are needed, we employ quantum Monte Carlo simulations, proving that the nontrivial topology of a thin shell allows superfluidity. 
Our work constitutes a reliable benchmark for the forthcoming scientific investigations with bubble traps. 
\end{abstract}

\maketitle

The recent advances in microgravity experiments with Bose-Einstein condensates have recently allowed us to extend the intrinsic limits of ground-based experiments and to realize exotic confining potentials for systems of ultracold atoms \cite{elliott,frye,vanzoest,meister,becker,aveline}. 
In particular, the seminal proposal by Zobay and Garraway to produce matter-wave condensate bubbles \cite{garraway1,garraway2,garraway3} is currently under investigation in NASA cold atom laboratory (CAL) on the international space station \cite{lundblad}. 
Experimentally, shell-shaped atomic traps are engineered by an adiabatic deformation of a conventional magnetic trap with a radiofrequency field. 
A quasi-two-dimensional hollow condensate can however be obtained only in microgravity conditions, since without any mechanism to compensate for gravity {the atoms fall to the bottom} of the trap \cite{colombe,demarco,harte}.

Spherically symmetric hollow condensates have a rich low-energy dynamical behavior \cite{lannert,padavic,sun,moller}, and the interplay of curvature, nontrivial contact interaction \cite{zhang}, and finite-size give rise to an interesting phase diagram in the thin-shell limit \cite{tononi,ovrut,bereta}. 
Moreover, it is expected that dipolar interactions induce anisotropic density profiles \cite{adhikari,diniz}, while for soft-core interactions a clusterization phenomenon is suggested \cite{prestipino}. 
In other hollow configurations as ring and toroidal traps \cite{ryu} a cooling quench may induce superfluid currents via the Kibble-Zurek mechanism\cite{kibble,zurek,corman}, but the adiabaticity requirements for bubble traps should prevent this phenomenon. 
All the recent papers deal with the simplified geometry of a spherical shell, and a complete physical description of the quantum statistics of an ellipsoidal shell is currently lacking. 

Inspired by the planned microgravity experiments \cite{lundblad}, we investigate the physics of a bosonic system of particles confined on an ellipsoidal shell. 
We calculate the critical temperature for Bose-Einstein condensation $T_{\text{BEC}}$ with a self-consistent Hartree-Fock (HF) theory \cite{giorgini,pitaevskii}. 
We find that, when the atoms are adiabatically transferred from the bare harmonic trap to the bubble trap, $T_{\text{BEC}}$ decreases up to a factor of $10$. 
This is partly due to the trap geometry, and partly due to a reduced maximal local density, which we estimate at $T_{\text{BEC}}$ with the HF theory, and at $T=0$ with the Gross-Pitaevskii equation \cite{gross,pitaevskiieq}. 
We also simulate the free expansion of the ellipsoidal shell: the peculiar topology of our system results in a new interference pattern, with qualitative differences from that of the harmonically trapped condensate. 
For temperatures lower than $5 \, \text{nK}$, 
mainly for $N \lesssim 5 \times 10^3$ particles, our semiclassical approach breaks down. 
To investigate quantitatively the coherence properties in this regime, we adopt a first-principle path integral Monte Carlo (PIMC) numerical approach \cite{ceperley}, which can accurately predict the physics of a mesoscopic trapped system \cite{cinti}. 

Our results are of great relevance for the forthcoming experiments with bubble traps, and for the future developments of microgravity physics, allowing a greater understanding of Bose-Einstein condensation and superfluidity in curved and compact manifolds. 

We consider a system of $^{87}$Rb atoms in the hyperfine state $\ket{F=2,m_F=2}$, confined in the three-dimensional harmonic potential $u(\vec{r}) = m \, (\omega_x^2 x^2 + \omega_y^2 y^2 + \omega_z^2 z^2)  /2$, where $m$ is the atomic mass, $\vec{\omega}=(\omega_x,\omega_y,\omega_z)$ are frequencies of the confinement, and $\vec{r}=(x,y,z)$. 
A shell-shaped condensate can be obtained by tuning a radio frequency magnetic field with a detuning $\Delta$, which must be much larger than the Rabi frequency $\Omega$ between the hyperfine levels \cite{lundblad}. 
If this dressing procedure is performed adiabatically, and under the hypothesis that any gravitational effect can be neglected \cite{note0}, the atoms will be confined by the bubble-trap potential \cite{garraway1} 
\begin{equation}
U(\vec{r}) = M_F \sqrt{[u(\vec{r})/2-\hbar\Delta]^2 + (\hbar\Omega)^2}, 
\label{bubble}
\end{equation}
where $M_F = 2$ now labels the higher dressed state with energy $U(\vec{r})$, and $\hbar$ is the Planck constant. 

Adopting Eq.~(\ref{bubble}) for the realistic experimental parameters of Ref. \cite{lundblad}, 
here we calculate the critical temperature $T_{\text{BEC}}$ of the transition between a noncondensed cloud and a Bose-Einstein condensate.
With a self-consistent Hartree-Fock theory \cite{giorgini}, the sum over all occupation numbers of thermal states, given by the Bose distribution at a fixed critical temperature $T_{\text{BEC}}$, is equal to the critical number of atoms $N$ at that temperature. 
In particular, the quasiparticle excitation spectrum appearing in the Bose distribution is treated semiclassically as $E(\vec{p},\vec{r}) = p^2/2 m + U(\vec{r}) + 2 g_0 n(\vec{r})$, where $n(\vec{r})$ is the number density, $p$ is the momentum of the excitation, $g_0=4\pi\hbar^2 a_{s} /m$ is the zero-range interaction strength, 
and $a_s = a_{\text{Rb}}$ is the three-dimensional $s$-wave scattering length of $^{87}\text{Rb}$. 
The external potential $U(\vec{r})$ is given by Eq.~(\ref{bubble}), in which we choose $\vec{\omega}/(2 \pi)=(30,100,100) \, \text{Hz}$, and set $\Omega/(2 \pi) = 5 \; \text{kHz}$ \cite{lundblad} throughout the Letter. 
Different trapping configurations, from thicker shells with a small size, 
to thinner ones with a larger size, can be obtained by choosing increasing values of the detunings $\Delta$, which can be experimentally tuned to engineer different traps. 

\begin{figure}[bt]
\includegraphics[scale=1.0]{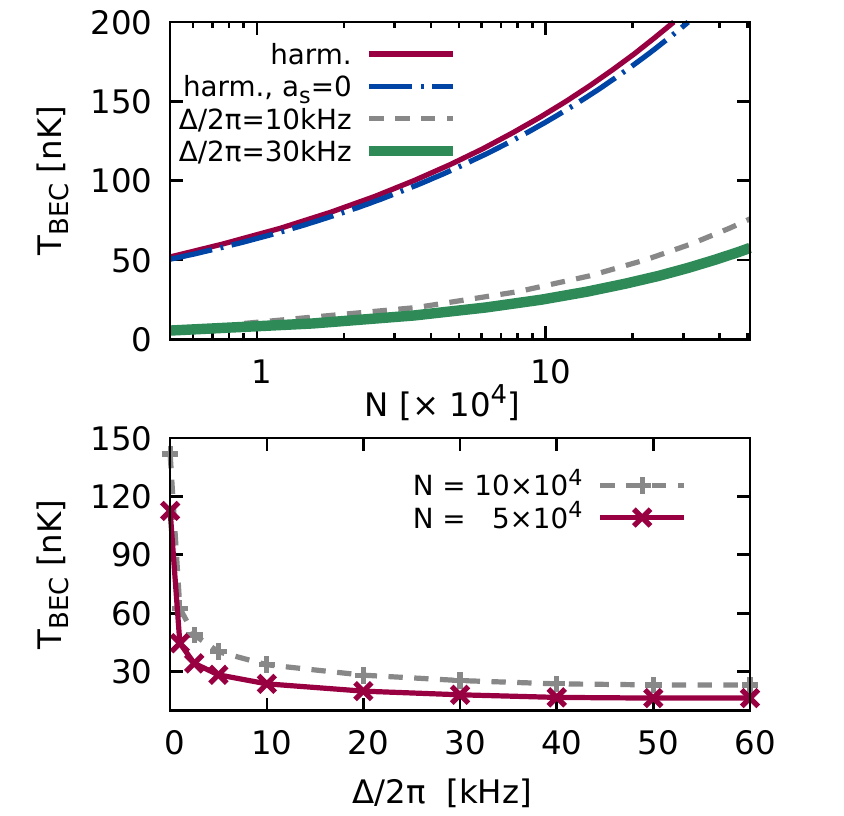}
\caption{
Critical temperature for Bose-Einstein
    condensation $T_{\text{BEC}}$ as a function of the number of particles $N$ (top)
    and detuning $\Delta$ (bottom).
Top: comparison for different external potentials: harmonic trap with $\vec{\omega}/(2 \pi)=(30,100,100) \, \text{Hz}$ (red thin line), noninteracting bosons in a harmonic trap \cite{dalfovo} (blue dot-dashed line), bubble trap with $\Delta/(2\pi)=10 \, \text{kHz}$ (grey dashed line), bubble trap with $\Delta/(2\pi)=30 \, \text{kHz}$ (green thick line). {Bottom: as} soon as $\Delta$ is nonzero $T_{\text{BEC}}$ decreases {partly} due to the reduced maximal local density, becoming essentially constant for large detunings \cite{sm}.
}
\label{fig1}
\end{figure}

Our results are summarized in Fig.~\ref{fig1}, in which $T_{\text{BEC}}$ is reported as a function of the particle number $N$ (top panel), and of the detuning $\Delta$ (bottom panel). 
The top panel clearly shows that quantum degeneracy is harder to reach in bubble traps than in conventional harmonic traps, with the critical temperature decreasing up to a factor of 10. 
Thus, even if the atomic cloud cools during the adiabatic deformation of the trap \cite{privcomm}, when the temperature in the predressed harmonic potential is not low enough an initial condensate may become a thermal cloud. 
We emphasize that, for a fixed particle number, the critical temperature of a thinner shell [$\Delta/(2\pi)=30 \, \text{kHz}$, green thick line] is slightly lower than the one of a thicker shell [$\Delta/(2\pi)=10 \, \text{kHz}$, grey dashed line]. 
A complementary picture is given by the bottom panel of Fig.~\ref{fig1}, where the critical temperature is shown to decrease quickly for an increasing detuning, with $\Delta=0$ corresponding to the bare harmonic trap (see also Ref. \cite{sm}). 
Further simulations show that, tuning the $s$-wave scattering length up to a factor 5 of the bare value for $^{87}\text{Rb}$, the critical number of particles decreases up to $20\%$ of Fig.~\ref{fig1} values. 
Moreover, we verified that this approach reproduces our previous results for a thin spherical shell \cite{tononi} as long as $N \gtrsim 10^5$ and $\Delta \gg \Omega \gg \omega$. 

With respect to current experiments, the previous results neglect the inhomogeneities of the potentials, which can be quantified as a $0.001 \, g$ tilt of the trap \cite{lundblad}, with $g$ the acceleration of gravity at the Earth level. {Under this hypothesis, we can also safely neglect the residual microgravitational corrections ($\sim 10^{-6}\,  g$) \cite{lundblad}.  
While the} inhomogeneities affect the atomic spatial distribution \cite{lundblad}, preventing a uniform condensation along the shell, we find that the critical temperatures of Fig.~\ref{fig1} are practically unchanged. 

The validity of our HF theory relies on the inequality $k_{B} T > \hbar \omega_{0}$, 
where $\omega_{0}$ is the typical frequency spacing between the levels of the system \cite{giorgini}. 
Following Ref.~\cite{sun}, we estimate $\omega_0 = \bar{\omega} \sqrt{2 \Delta/\Omega}$, with $\bar{\omega} = (\omega_x \omega_y \omega_z)^{1/3}$ 
the geometric average of the harmonic trap frequencies, 
so that the minimum temperatures at which the semiclassical approximation 
is expected to hold are $\hbar \omega_0 /k_{B} \approx 5 \, \text{nK}$. 
Our theory is reliable over this critical temperature, which corresponds to $N \gtrsim 5 \times 10^3$. 

\begin{figure}[b]
\includegraphics[scale=1.0]{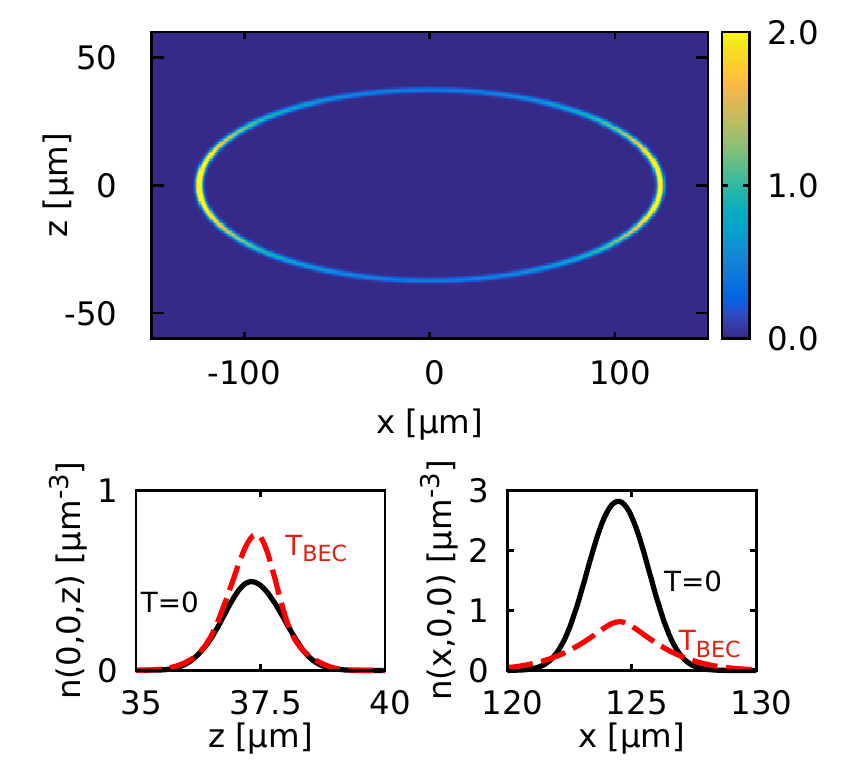}
\caption{\label{fig2}Top: contour plot of the density in the $xz$ plane (colorbox units in $\mu\text{m}^{-3}$), obtained solving Eq.~(\ref{GPE}), for $\vec{\omega}/(2 \pi)=(30,100,100) \, \text{Hz}$, $\Delta/(2 \pi) = 30 \; \text{kHz}$, $\Omega/(2 \pi) = 5 \; \text{kHz}$, and $N=57100$. 
Bottom: one-dimensional sections of the density at $T=0$ (from the GPE) and at $T_{\text{BEC}}$ (from Hartree-Fock theory). Note that at $T=0$ the condensate is concentrated on the shell vertices, while the thermal cloud at $T_{\text{BEC}}$ is uniformly distributed. 
}
\end{figure}

For a fixed particle number $N$, at temperatures sufficiently lower than those identified in Fig.~\ref{fig1}, all the particles of this weakly interacting system can be approximately thought to be in the same single-particle state. 
In this zero-temperature fully condensate regime, the macroscopic wave function of the system $\psi(\vec{r},t)$ satisfies the Gross-Pitaevski equation (GPE) \cite{gross,pitaevskiieq} 
\begin{equation}
i \hbar \frac{\partial \psi(\vec{r},t)}{\partial t} = \bigg[ - \frac{\hbar^2 \nabla^2}{2m} + U(\vec{r}) + g_{0} |\psi(\vec{r},t)|^2  \bigg] \psi(\vec{r},t).
\label{GPE}
\end{equation}
The stationary solution of Eq.~(\ref{GPE}) gives the condensate density at zero temperature, i.e. $n(\vec{r}) = |\psi(\vec{r})|^2$. 
In particular, by using an imaginary-time propagation algorithm \cite{chiofalo}, here we solve the GPE for $N=57100$ bosons trapped in the external potential of Eq.~(\ref{bubble}) with $\Delta/(2 \pi) = 30 \; \text{kHz}$, and $\Omega/(2 \pi) = 5 \; \text{kHz}$ \cite{N}. 
In the top panel of Fig.~\ref{fig2} we plot a two-dimensional section of the condensate density $n(\vec{r})$, cut along the $xz$ plane. 
For simplicity, we avoid showing the density distribution along the $xy$ plane, due to the trivial axial symmetry of the confinement. 
We find that the particles are not uniformly distributed on the shell, but accumulate on the ellipsoid lobes, where the local ellipsoidal trapping is weaker. 
This nonuniform particle distribution across the shell can be also seen in the bottom panels of Fig.~\ref{fig2}, in which we plot the one-dimensional cuts of the condensate density $n(\vec{r})$ along the $x$ and the $z$ direction ($T=0$ label) \cite{note}. 
It is quite interesting to compare the condensate distribution with the thermal density at the critical temperature $T_{\text{BEC}}$, obtained from the HF theory ($T_{\text{BEC}}$ label).
Similarly to harmonically trapped gases \cite{giorgini}, while the condensate density at $T=0$ is more localized in the vertices, the thermal cloud is broader and practically uniform: this crucial difference can be used as a first experimental check of the temperature of the system.  
At the same time, given the current status of microgravity experiments, the observation of these effects requires a precise control of the inhomogeneities of the radio frequency field, to get a full and symmetric coverage of the shell. This is the object of ongoing experimental efforts on CAL \cite{lundblad}, towards the next generation of experiments on BECCAL \cite{frye}.

\begin{figure}[b]
\includegraphics[scale=1.0]{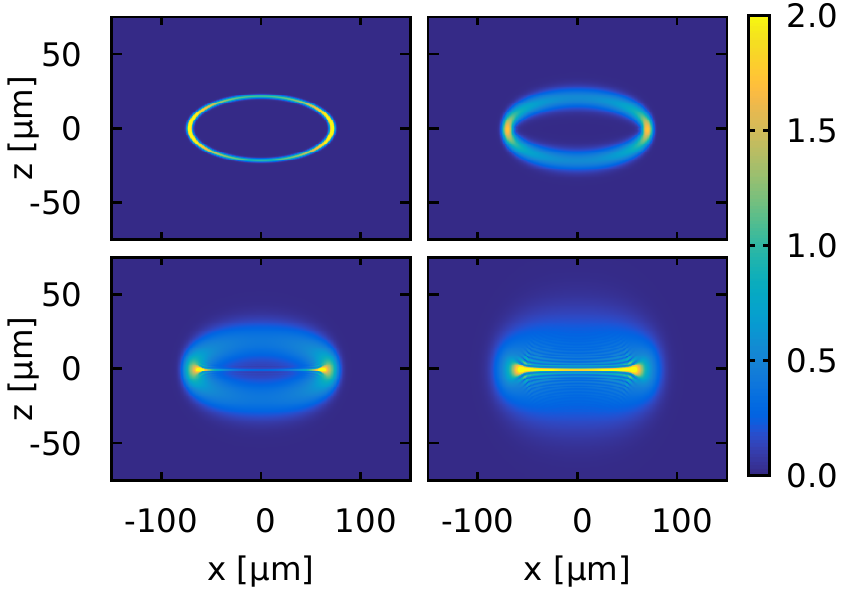}
\caption{\label{fig3}Free expansion of the condensate shell, initially in the ground state for the bubble-trap potential of Eq.~(\ref{bubble}) with $\Delta/(2 \pi) = 10 \; \text{kHz}$, and the other parameters as in Fig.~\ref{fig2}. 
From left to right and from top to bottom, the condensate slices along the xz plane are taken at the times: $0 \, \text{ms}$, $4.5 \, \text{ms}$, $9 \, \text{ms}$, and $18 \, \text{ms}$. The hollow condensate expands both outwards and inwards, showing a qualitatively different interference pattern with respect to that of harmonic traps \cite{sm,salasnich}.
}
\end{figure}

To analyze more deeply the physics of bubble-trapped condensates we now study the dynamics of the system, by solving numerically Eq.~(\ref{GPE}) \cite{numericalrecipes}. 
The peculiar signatures of a hollow Bose-Einstein condensate clearly emerge in the free expansion of the system. 
In this case, without any magnetic confinement, the hyperfine splitting of the atomic energy levels is absent, and the simulation of a single GPE is sufficient \cite{note2}. 
Starting from the stationary solution of Eq.~(\ref{GPE}) for $\Delta/(2 \pi) = 10 \; \text{kHz}$ and $N=57100$ bosons, we suddenly remove the trapping potential $U(\vec{r})$ at the time $t= 0 \, \text{ms}$. During the dynamics of the system we take three snapshots of the density in the xz plane, for $4.5$, $9$, and $18 \, \text{ms}$. 
The last panel of Fig.~\ref{fig3} depicts the interesting interference pattern obtained when the matter-wave self interferes at the center of the trap. 
As a qualitative difference with respect to  harmonically trapped condensates \cite{sm}, here we observe the appearance of a central density peak around the final time of $18 \, \text{ms}$. 
Since the free expansion of the condensate cloud takes place in a $\sim 10 \, \text{ms}$ time, and the main interference peak has a width of approximately $4 \, \mu\text{m}$ \cite{sm}, this phenomenon is easily observable in current microgravity experiments. 

For a low number of particles, the Hartree-Fock theory is not expected to describe 
accurately the physics of the system. 
In this regime, we describe the coherence properties through a continuous-space worm algorithm PIMC numerical simulation \cite{boninsegni1,boninsegni2}. 
This technique allows us to simulate the exact dynamics of the system, described 
by the general Hamiltonian 
\begin{equation}
{\cal H} = - \frac{1}{2} \sum_{i=1}^{N} \nabla^2 + 
\sum_{i=1}^{N} U(\vec{r}_i) + \sum_{i<j}^{N} v(|\vec{r}_i-\vec{r}_j|), 
\label{hamiltonian}
\end{equation}
in which the particles are interacting with the hardcore potential $v(|\vec{r}_i-\vec{r}_j|) = \infty$ for $|\vec{r}_i-\vec{r}_j| < r_0$, with $r_0$ the hardcore potential range, and $0$ otherwise.
We stress that in the Hamiltonian (\ref{hamiltonian}) we have rescaled all the energies with $\hbar^2 /(m r_0^2)$, the typical energy of the two-body interaction. 
In particular, for our interaction potential the range $r_0$ can be identified with the three-dimensional $s$-wave scattering length $a_s$ \cite{stoof}. 

In order to observe superfluidity in dilute systems of few particles ($N \lesssim 10^3$), it is crucial that the interaction between bosons is strong enough. In this case, the critical rotational frequency of the trap $\Omega_c \approx g_0 n/\hbar$ under which superfluidity can occur will be sufficiently large, and the system will be in the Thomas-Fermi regime.
Clearly, the experimental observation of superfluidity in a mesoscopic system is a trade-off between the necessity of having a long enough lifetime of the condensate to perform measures, and the intrinsic limits of the cooling apparatus. 
To discuss temperature regimes that are relevant for the current experimental capabilities ($T \gtrsim \text{nK}$), we model a Bose gas in which the scattering length is tuned with 
a Feshbach resonance to the value $a_s = c \; a_{\text{Rb}}$, 
where $a_{\text{Rb}}$ is the bare scattering length of $^{87}\text{Rb}$ \cite{egorov}. 
The scattering length of $^{87}\text{Rb}$ can be tuned 
with the $1007 \, G$ Feshbach resonance \cite{marte,smirne} up to $c \approx 10$, at least without 	reducing significantly the number of trapped atoms. For a mesoscopic system, we suggest that $c$ can reasonably be tuned up to $25 \div 50$.

\begin{figure}[t]
\begin{center}
\resizebox{0.99\columnwidth}{!}{\includegraphics{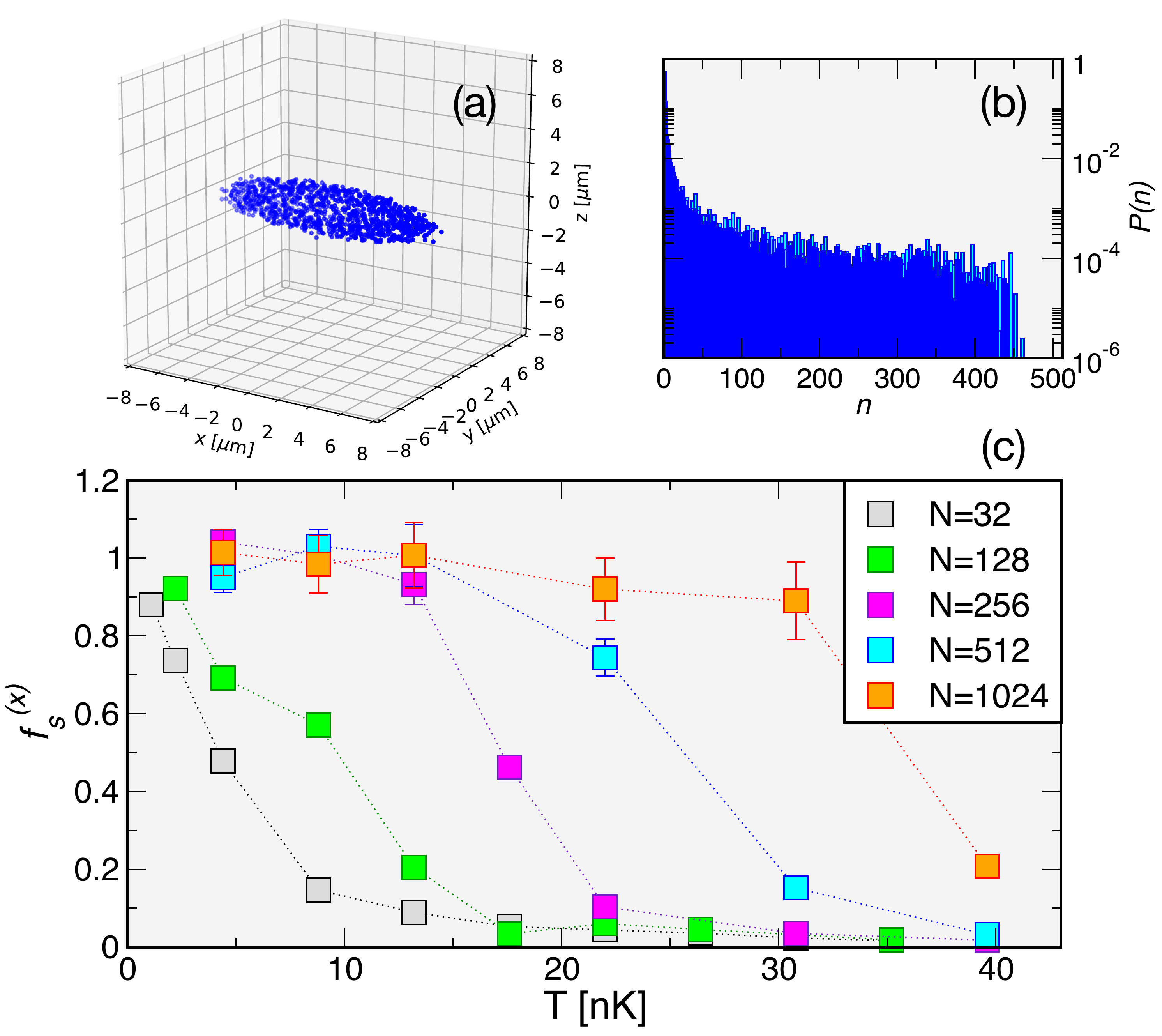}}
\caption{\label{fig4}Results of the Monte Carlo simulations, in which we employ the bubble-trap potential of Eq.~(\ref{bubble}) with $\vec{\omega}/(2 \pi)=(0.2,1,1) \, \text{kHz}$, $\Delta/(2 \pi) = 10 \; \text{kHz}$, $\Omega/(2 \pi) = 5 \; \text{kHz}$, and $a_s=50 \, a_{\text{Rb}}$. 
In (a) we represent the real space projection of the wordlines for $N=512$ bosons at $4.4 \, \text{nK}$. 
The superfluid character of this configuration is proven in panel (b), by the fat-tailed distribution $P(n)$ of the $n$-particles permutation cycles, with $1 \leq n \leq N$. 
We summarize our simulations in panel (c), showing the superfluid fraction of the system $f_s^{(x)}$ as a function of the temperature $T$.}
\end{center}
\end{figure}

The results of our simulations are shown in Fig.~\ref{fig4}. In particular, Fig.~\ref{fig4}(a) depicts an instantaneous configuration of $N=512$ bosons at a temperature of $T=4.4\, \text{nK}$, featuring the projection of world lines onto real space. 
These projections have the important insight to be the closest representation of the square of the many-body wave function~\cite{ceperley,PhysRevLett.119.215302}. 
As a result, Fig.~\ref{fig4}(a) displays an evident paths overlapping which implies exchanges among delocalized particles and hence global superfluidity. 
This claim finds agreement with Fig.~\ref{fig4}(b) where we report the relative probability $P(n)$ that $n$-particles exchange within the bubble-trap confinement $U(\vec{r})$. 
Note that $P(n)$ is nonzero on an extended region of $n$, concerning long permutation cycles (exchanges) of the order of $n\lesssim N$. 

Let us now quantitatively discuss the phenomenon of superfluidity in this strongly-interacting hollow gas. 
In a confined system, the superfluid fraction can be calculated as the ratio of the non-classical inertial moment $I_i$ and the classical one $I_{i}^{\text{cl}}$, the index $i$ being one of the main axes along the directions $x$, $y$, and $z$. 
Thus, the estimator of the superfluid fraction $f_s^{(i)}$ is given by \cite{kwon2006,Jain2011,zeng} $f_s^{(i)}=4m^2 \left\langle A_i^2\right\rangle/(\hbar^2 \beta\left\langle I_{i}^{\text{cl}}\right\rangle)$, where $\beta=1/k_BT$, while $\langle  \cdots \rangle$ stands for the thermal average, and $A_i$ underlies the world-line area of closed particle trajectories projected on its corresponding perpendicular plane \cite{kwon2006,Jain2011,zeng}. 
The superfluid fraction $f_s^{(x)}$ is reported in Fig.~\ref{fig4}(c) as a function of the temperature $T$, showing the results of the sampling for $N$ ranging from 32 to 1024 bosons. 
We stress that, when increasing the number of bosons $N$, the coherence effects are enhanced, and a sizeable superfluid fraction is reached at higher temperatures. 
Regarding $f_s^{(y)}$ and $f_s^{(z)}$, we find that they result systematically lower than $f_s^{(x)}$ by a factor of 5 \cite{sm}. 
This result implies an anisotropic second sound velocity, which in a two-dimensional weakly interacting bosonic system goes as $c_2^{(i)} \propto (f_s^{(i)})^{1/2}$ \cite{stringari2014}. 
Experimentally, a density perturbation in a sufficiently large and flat shell will then show that $c_2^{(x)}>c_2^{(z)}$ \cite{sm}. 
Similarly to what we have deduced with the {HF} theory, we have verified that for fixed $N$ and $a_s$, the superfluid fraction is lower for thinner and larger ellipsoidal shells, in which the collective behavior is suppressed. 
Moreover, we have also verified that the typical temperature range at which $f_s^{(x)}$ becomes significant in a bubble-trap are up to a factor of 5 lower than the ones for a harmonically trapped gas. 
Finally, since we are simulating a finite-size small system, there is not a finite temperature at which the superfluid fraction vanishes, but increasing $N$ the transition will get sharper and the residual $f_s^{(x)}$ will tend to zero. 
All these observations clearly show that, despite the topology of a thin shell-shaped condensate is different from the one of the 2D flat plane, the system is superfluid. 

To conclude, we have calculated the critical temperature for Bose-Einstein condensation of a bosonic system of atoms confined on a shell-shaped potential, finding that, with respect to the bare harmonic trap, the critical temperature is significantly lower. 
We have then simulated the Gross-Pitaevskii equation with the realistic external potential parameters to describe the ground state and the free expansion of the system, observing an interesting self-interference pattern during the hole filling. 
Finally, we have shown that for a mesoscopic number of particles in a regime of strong interactions the thin atomic shell is superfluid for experimentally accessible temperature regimes. 
Our findings will be of great interest for modeling and understanding 
the ongoing experiments with microgravity Bose-Einstein condensates. 

\begin{acknowledgments}
A. T. thanks N. Lundblad for insightful discussions, and acknowledges useful discussions with B. Garraway, and the participants of the ``BECCAL Brainstorming Workshop,'' held in Ulm in December 2019. A. T. and L. S. thank F. Ancilotto for enlightening comments and suggestions, and A. Trovato and A. Cappellaro for useful suggestions. 
L.S. acknowledges the BIRD project ``Time-dependent density functional theory of quantum atomic mixtures'' of the University of Padova for partial support.
CloudVeneto is acknowledged for the use of computing and storage facilities.
\end{acknowledgments}

\section*{Supplemental Material for ``Quantum bubbles in microgravity''}

\renewcommand\thefigure{S\arabic{figure}}
\setcounter{figure}{0}
\renewcommand\theequation{S\arabic{equation}}
\setcounter{equation}{0}

\begin{center}
\textbf{On BEC critical temperature}
\end{center}
For a fixed Rabi frequency $\Omega$, and for fixed harmonic trap frequencies $\vec{\omega}$, the bubble-trap potential $U(\vec{r})$ depends only on the detuning $\Delta$. 
Increasing the value of $\Delta$, the size of the condensate shell increases, and the trap becomes tighter in the radial direction. 
Correspondingly, the critical temperature drops quickly to an almost constant value, as can be seen in Fig.~1 of the main paper. 
A phenomenological fitting formula of the critical temperature for $\Delta/(2\pi) \gtrsim 5 \, \text{kHz}$ is given by 
\begin{equation}
\nonumber
T_{\text{BEC}} \approx 15 \, \exp\bigg(\frac{6.08}{4.65+\Delta}\bigg)
\end{equation}
for $N=5 \times 10^4$ particles, and by 
\begin{equation}
\nonumber
T_{\text{BEC}} \approx 21.17 \, \exp\bigg(\frac{6.45}{5+\Delta}\bigg)
\end{equation}
for $N=10 \times 10^4$ particles. 

\begin{center}
\textbf{Free expansion: comparison with the harmonic trap.}
\end{center}

\begin{figure}[hbtp]
\includegraphics[scale=1.0]{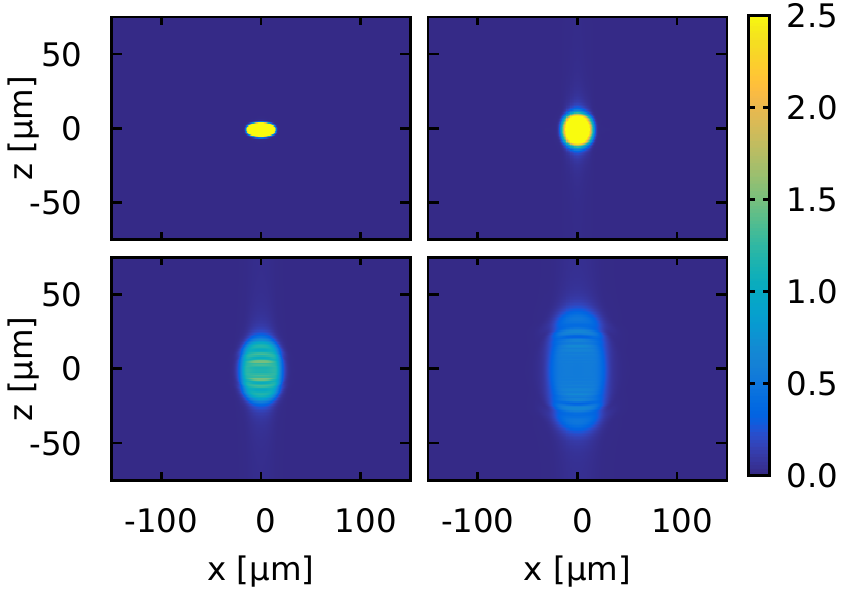}
\caption{Density cuts in the xz plane during the free expansion of the condensate. The atoms were initially in the ground state of an harmonic trap with frequencies $\vec{\omega}/(2 \pi)=(30,100,100) \, \text{Hz}$, while, in Fig.~3 of the main text, the atoms were initially confined in a bubble trap with the same bare harmonic frequencies. As in the main text, here we use $N=57100$, and consider interacting $^{87}$Rb atoms. 
}
\label{figsm1}
\end{figure}

In Fig.~\ref{figsm1} we show four density cuts in the xz plane taken during the free expansion of harmonically trapped gas, at the times $0 \, \text{ms}$, $4.5 \, \text{ms}$, $9 \, \text{ms}$, and $18 \, \text{ms}$. 
Note that the frequencies of the bare harmonic confinement are the same of those used for the bubble trap, namely $\vec{\omega}/(2 \pi)=(30,100,100) \, \text{Hz}$. 
The self-interference of the harmonically trapped condensate shows qualitative differences with the free expansion of the condensate shell, as can be seen by comparing Fig.~\ref{figsm1} with Fig.~3 of the main paper.

\begin{figure}[hbtp]
\includegraphics[scale=1.0]{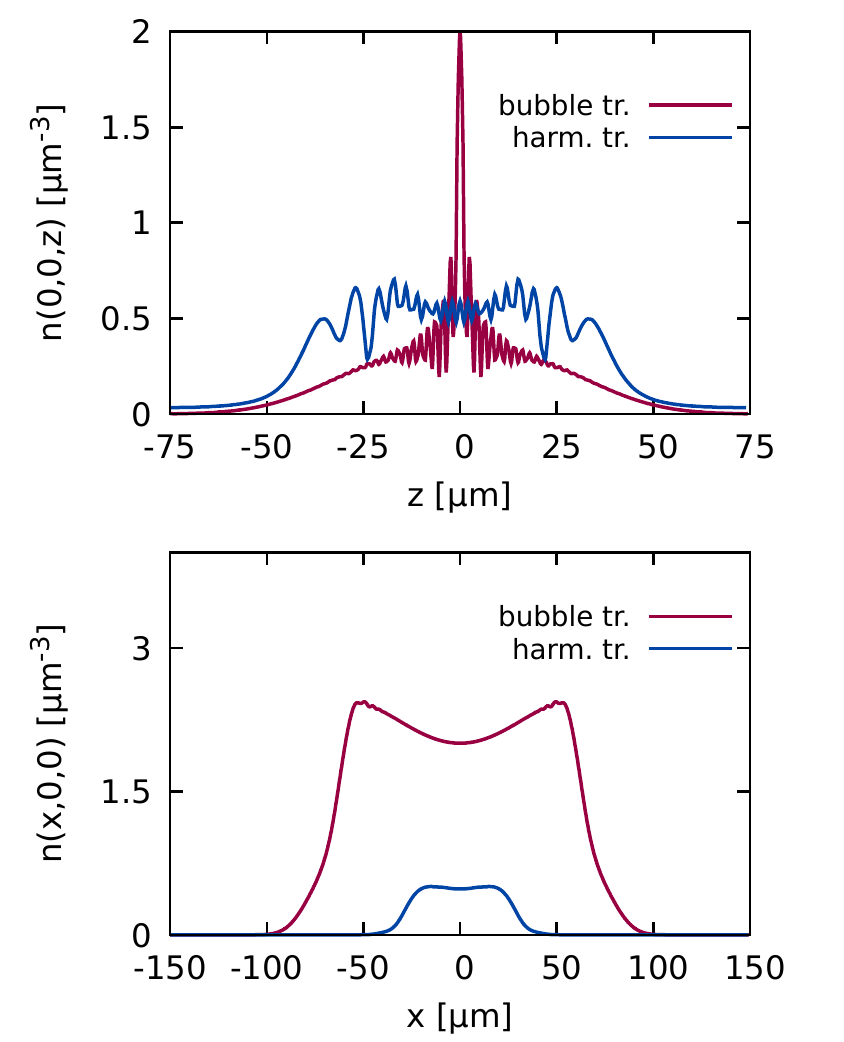}
\caption{Condensate density along the axes $x$, and $z$, taken $18 \, \text{ms}$ after releasing the trap.  For the atoms initially confined in the bubble trap, the central density peak shown in the top panel has a width of $4 \, \mu\text{m}$. In both simulations the number of particles is $N=57100$.  
}
\label{figsm2}
\end{figure}

The different expansion of the atoms in these configurations can also be seen from the density cuts along the main system axes. In Fig.~\ref{figsm2} we compare the one-dimensional density profiles of the bubble trap (red line) and of the harmonic trap (blue line). The hollow configuration of the shell allows a quick expansion of the condensate in the inward direction, with the formation of a central density peak along the $x$ axis.

\newpage
\begin{center}
\textbf{Anisotropic superfluid fraction}
\end{center}

\begin{figure}[hbtp]
\includegraphics[scale=0.4]{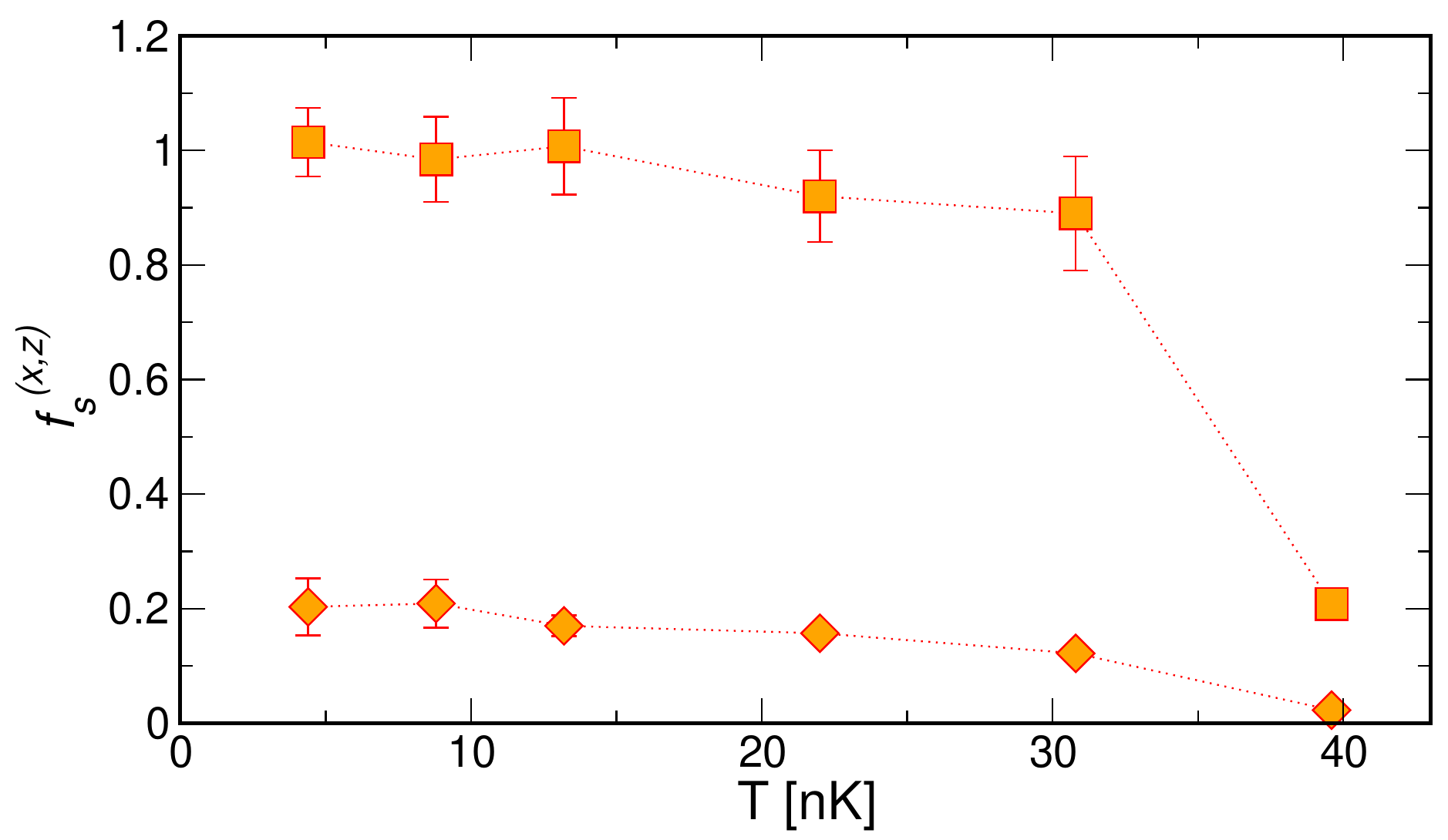}
\caption{Anisotropic superfluid fraction for a bosonic system of $N=1024$ atoms, obtained with PIMC simulations. 
Compatibly with the statistical error bars, we find that $f_s^{(x)}$ (square points) is compatible to $1$ up to temperatures of $\sim 30 \, \text{nK}$, and it is larger than $f_s^{(z)}$ (diamond points) up to a factor of $\sim 5$.
}
\label{figsm3}
\end{figure}

Our Path Integral Monte Carlo simulations show that the superfluid fraction of the bubble trap is anisotropic. 
Indeed, due to the rotational symmetry breaking by the anisotropic shape of the bubble trap, one finds that $f_s^{(x)} > f_s^{(z)}$. 
In Fig. \ref{figsm3} we compare $f_s^{(x)}$ and $f_s^{(z)}$ for $N=1024$ bosons. 

In two-dimensional weakly-interacting bosonic gases, a density probe excites mainly the second sound, with a velocity $c_s$ proportional to the square root of the superfluid fraction \cite{ozawa}. 
Assuming that the transverse modes of the shell are not excited, one can effectively describe the condensate shell with the two-dimensional density $n(\theta,\phi,t)$. Here $\theta \in [0,\pi]$, and $\phi \in [0,2\pi]$ are the elliptical coordinates along the shell.  
To excite the second sound in the azimuthal direction $\phi$ we consider a density probe, for instance a potential step suddenly turned on, elongated in the direction $\theta$. 
For a fixed value of $\theta=\bar{\theta}$, we can  decompose the density as \cite{noncollisional}
\begin{equation}
\nonumber
n(\bar{\theta},\phi,t) = n_0(\bar{\theta}) + \sum_{m_l >1} A_{m_l} (t) \, \text{e}^{i m_l \phi}. 
\end{equation}
Assuming a weak perturbation, only $A_1$ will be excited: by measuring the frequency $\omega$ of the oscillations of $A_1(t)$ one can calculate the sound velocity as $c=\omega b$, with $b$ the minor axis of the prolate shell. 
Note that the second sound will be excited only in the collisional regime of $\omega < \Gamma_{coll}$ \cite{cappellaro}, with $\Gamma_{coll} \approx g_0 n / \hbar \approx 50 \, \text{Hz}$ in our system. 
A similar calculation applies to the second sound propagation along the $\theta$ direction, excited by a density probe along $\phi$. Due to the anisotropic superfluid fraction we predict an anisotropic sound velocity, with $c_s^{(x)} > c_s^{(z)}$. 

\begin{thebibliography}{50}
\bibitem{note0} {Performing the experiments in microgravity allows us to neglect the additional $mgz$ contribution to the external potential which affects ground-based experiments \cite{colombe}. }
\bibitem{N} We employ these parameters to be to be as close as possible to the experiments, and to the previous literature on this topic, see Ref. \cite{lundblad}.
\bibitem{note} The level of discretization for this two-dimensional condensate does not allow to observe the radial discontinuities in the density profile of Ref. \cite{giorgini}.
\bibitem{note2} In principle the $s$-wave scattering length for the atoms in the various hyperfine levels can be different, but due to the low atomic density of approximately $10^{12} \, \text{cm}^{-3}$ the use of $g_0$ is a good approximation.
\end{thebibliography}
\end{document}